\documentclass[superscriptaddress,pra,twocolumn,showpacs,preprintnumbers,altaffilletter,aps,10pt,nofootinbib]{revtex4-1}
\usepackage{amssymb}
\usepackage{graphicx}
\usepackage{setspace}
\usepackage[caption=false]{subfig}
\usepackage{lipsum}
\usepackage{multirow}
\usepackage{siunitx}
\usepackage{lineno}
\usepackage{datetime}
\usepackage{xspace}
\usepackage{hyphenat}
\usepackage{bbold}
\usepackage{mathtools} 
\usepackage{easyReview}
\usepackage{ulem}
\usepackage{enumitem}
\usepackage{todonotes}
\usepackage{siunitx}
\usepackage{upgreek}
\usepackage{hyperref}
\usepackage{color}
\definecolor{linkcol}{rgb}{0,0,0.4} 
\definecolor{citecol}{rgb}{0.5,0,0} 
\hypersetup
{
pdftitle="Determination of diffusion coefficients of mercury atoms in various gases from longitudinal spin relaxation in magnetic gradients",
unicode=false,          
pdftoolbar=true,        
pdfmenubar=true,        
pdffitwindow=true,     
pdfstartview={FitH},    
colorlinks=true,       
linkcolor=blue,          
citecolor=blue,        
}

\newcommand{\ds}{\displaystyle}
\newcommand{\cmsps}{\ensuremath{\mathrm{cm^2/s}}\xspace}
\newcommand{\muT}{\ensuremath{\mathrm{\upmu T}}\xspace}

\newcommand{\mbar}{\ensuremath{\mathrm{mbar}}\xspace}
\newcommand{\he}{\ensuremath{\mathrm{He}}\xspace}
\newcommand{\hg}{\ensuremath{\mathrm{Hg}}\xspace}

\begin{document}
\title{Determination of diffusion coefficients of mercury atoms in various gases from longitudinal spin relaxation in magnetic gradients} 
\author{B.~Cl\'ement} 
\affiliation{Univ. Grenoble Alpes, CNRS, Grenoble INP, LPSC-IN2P3, 38000 Grenoble, France}
\author{M.~Guigue} 
\affiliation{Sorbonne Universit\'e, Universit\'e Paris Cit\'e, CNRS/IN2P3, Laboratoire
de Physique Nucl\'eaire et de Hautes \'Energies (LPNHE), 75005 Paris, France}
\author{A.~Leredde} 
\affiliation{Univ. Grenoble Alpes, CNRS, Grenoble INP, LPSC-IN2P3, 38000 Grenoble, France}
\author{G.~Pignol} 
\affiliation{Univ. Grenoble Alpes, CNRS, Grenoble INP, LPSC-IN2P3, 38000 Grenoble, France}
\author{D.~Rebreyend} 
\affiliation{Univ. Grenoble Alpes, CNRS, Grenoble INP, LPSC-IN2P3, 38000 Grenoble, France}
\author{S.~Roccia} 
\affiliation{Univ. Grenoble Alpes, CNRS, Grenoble INP, LPSC-IN2P3, 38000 Grenoble, France}
\author{S.~Touati} 
\affiliation{Univ. Grenoble Alpes, CNRS, Grenoble INP, LPSC-IN2P3, 38000 Grenoble, France}

\begin{abstract}
We present a novel method to measure the binary diffusion coefficient of mercury atoms in a gas at room temperature and low pressure. 
It is based on the measurement of the longitudinal spin relaxation of optically pumped mercury-199 atoms in a magnetic field gradient. 
We provide a consistent set of diffusion coefficients for helium-3, helium-4, argon, krypton, xenon, nitrogen, carbon dioxide, oxygen, and air. 
\end{abstract}
\maketitle

\section{Introduction}


Nuclear Magnetic Resonance (NMR) provides a versatile and powerful probe of diffusion phenomena. 
In their seminal work in 1954, Carr and Purcell \cite{Carr1954} determined the self-diffusion coefficient of ordinary water by measuring the decay of transverse spin-polarization in a magnetic field gradient. 
Variants of this technique, such as Pulsed Gradient Spin Echo, are now routinely used to measure diffusion coefficients in liquids, which are of the order of $10^{-5}~\cmsps$. 
In gases, although diffusion NMR experiments are less ubiquitous, diverse techniques have been employed with hyperpolarized noble gases to measure the diffusion coefficient of helium-3 \cite{Barbe1974,Bock1997,Hayden2004,Acosta2006} or xenon-129 \cite{Hasson1990,Liu2016a} in different gases. 
Compared to the case of liquids, the diffusion is much faster in gases, of the order of $0.1~\cmsps$ at standard temperature and pressure. 

In this article we report on a study of the diffusion of mercury atoms in various gases at room temperature (20$^{\circ}$C). 
In the present experiment, the mercury is polarized in a very low field with optical pumping in a cell filled with mercury-199 at a partial pressure of about $10^{-5}~\mbar$. 
In addition the cell contains a buffer gas at a pressure in the range $0.5 - 5~\mbar$. 
In such a situation, the mercury atoms diffuse in the cell in a very short time, about $10~\mathrm{ms}$. 
Contrary to the case of liquids, for which the diffusion time across a sample of size 1~cm is about $10^5~\mathrm{s}$, the mercury atoms  explore the entire cell many times during the measurement duration of a few seconds. 
We have developed a method to determine the diffusion coefficient in this ``motion narrowing" regime, based on the measurement of the longitudinal relaxation rate (rather than the decay of the transverse polarization in most NMR diffusion experiments) in an applied magnetic field gradient. 

The motivation for this work comes from the prospects for a precision measurement of the gyromagnetic ratio of mercury-199 using helium-3 as a comagnetometer in the same cell. 
The helium-3 will be polarized by metastable-exchange optical pumping and its precession will be read out with induction coils. 
This technique is adequate for helium-3 operating in the millibar pressure range. 
Then, the presence of the helium-3 buffer gas will increase the time it takes for mercury atoms to average the field over the cell. 
In such a situation the residual magnetic field gradients could significantly shorten the decay time of the precession signal, and therefore affect the precision of the measurement of the Larmor precession frequency. 
In order to quantify the requirement on the acceptable field uniformity in the experiment, a dedicated measurement of the diffusion coefficient of mercury in helium-3 is desired. 

The rest of the article is organized as follows. 
In section \ref{sec:diffusion-relaxation} we review the treatment of longitudinal relaxation of the spins (i.e. the decay of the longitudinal component of the polarization) due to field gradients. 
The experimental apparatus and the analysis procedure are described in sections \ref{sec:experiment} and \ref{sec:analysis}. 
In section \ref{sec:results} we present our results for the binary diffusion coefficient of mercury in  various gases (helium-3, helium-4, argon, krypton, xenon, nitrogen, carbon dioxide, oxygen, and air), and compare them with previous measurements using non-NMR methods (see \cite{Marrero1972} for a comprehensive review) and the predictions based on Chapman-Enskog theory. 

\section{Longitudinal relaxation due to field gradient}
\label{sec:diffusion-relaxation}

Consider an ensemble of polarized spin $1/2$ atoms confined in a cell exposed to a non-uniform magnetic field. 
Due to their random motion through the cell, each atom effectively sees a fluctuating magnetic field, resulting in depolarization of the ensemble  \cite{Bloembergen1948,Bouchiat1960, Gamblin1965, Schearer1965,Redfield1965,Cates1988,McGregor1990,Golub2010,Pignol2015b}. 
In the motional narrowing case, i.e. when the depolarization is slow compared with the correlation time $\tau_D$ of the fluctuating field, the longitudinal and transverse polarization decay exponentially with relaxation rates $\Gamma_1$ and $\Gamma_2$  given by Redfield's theory \cite{Redfield1965}. 
In sufficiently high holding magnetic field $B$, that we take along the $x$ axis, the longitudinal relaxation rate due to the non-uniformity of the magnetic field is given by: 
\begin{equation}\label{eq:gamma1-adiab-diff}
    \Gamma_1 = D\frac{G^2}{B^2}, 
\end{equation}
where $D$ is the diffusion coefficient and $G$ is the gradient of the transverse field components, more precisely $G^2 = \langle (\vec\nabla B_y)^2 + (\vec\nabla B_z)^2 \rangle$ \cite{Guigue2014d}. 

\begin{figure}
    \includegraphics[width=0.45\textwidth]{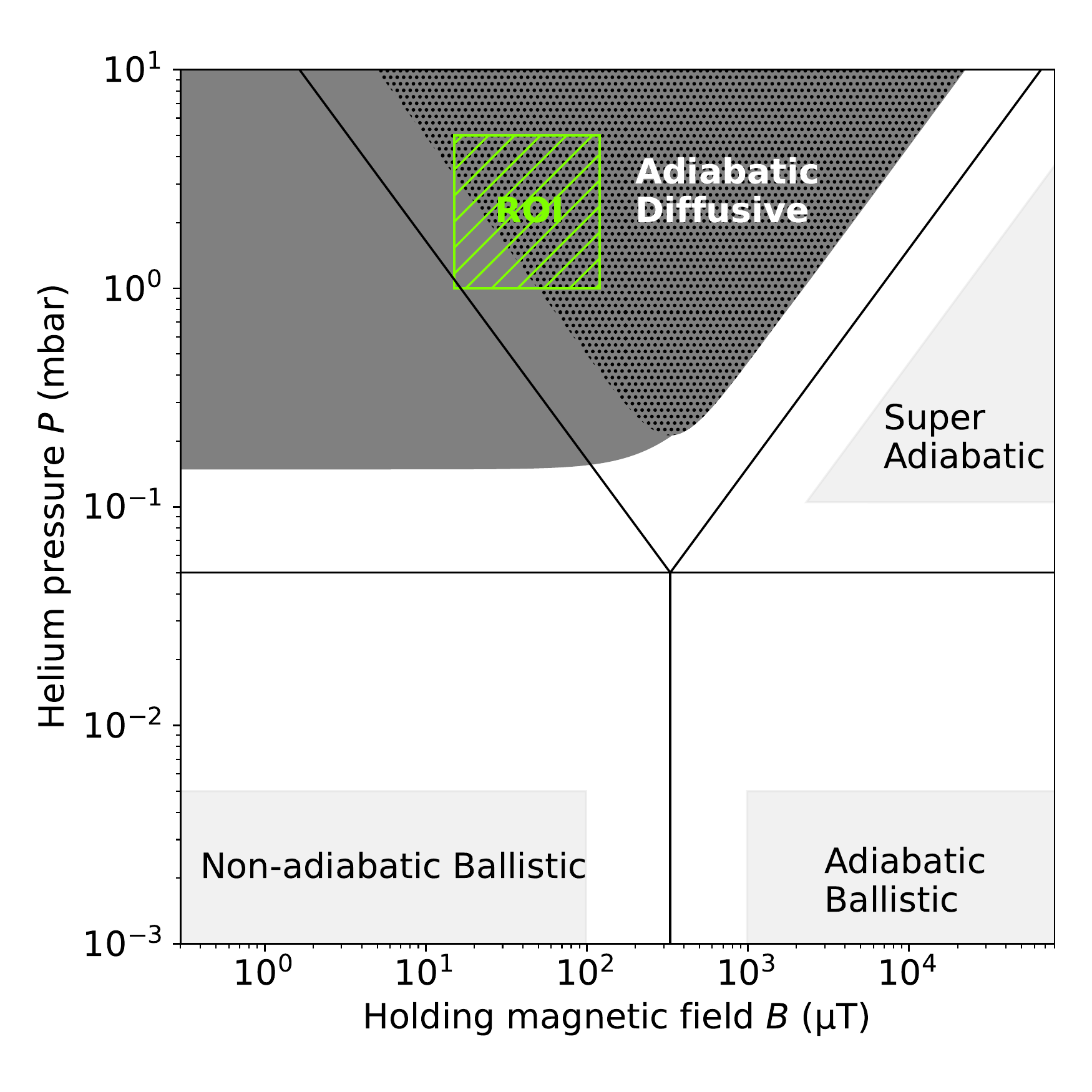}
    \caption{\label{fig:classification}
    Different regimes for the longitudinal relaxation of mercury atoms in a helium buffer gas.
    The horizontal line corresponds to the separation between diffusive and ballistic regimes. The upper gray zone corresponds to the validity domain of Eq. \eqref{eq:gamma1-final}, including the dotted region corresponding to the validity domain of Eq. \eqref{eq:gamma1-adiab-diff}. 
    The chartreuse dashed region, marked ROI,  corresponds to the parameter space explored experimentally in this study.}
\end{figure}

Let us now discuss in detail the validity condition of Eq. \eqref{eq:gamma1-adiab-diff}. 
For that, we consider for definiteness a cell of cylindrical shape, of radius $R=2$~cm and length $L=10$~cm, filled with a mixture of mercury and helium-4 atoms at room temperature. 
The helium pressure $P$ and the value of the holding magnetic field $B$ define different regimes  for the longitudinal relaxation, shown in Fig.~ \ref{fig:classification}. 

First let us consider the pressure. Without helium, mercury atoms would have a ballistic motion in the cell,  with a time between two wall collisions set by $2R/v \approx 0.4$~ms, $v=11$~cm/ms being the thermal velocity of mercury atoms at room temperature. 
Note that atomic collisions between two mercury atoms are much less frequent than wall collisions, because the pressure of mercury atoms in the cell needs to be very low (about $10^{-5}$~mbar in our experiment and in any case below the vapor pressure of $2 \times 10^{-3}$~mbar at room temperature). 
Now, for sufficiently large pressure $P$ of helium in the cell, the mercury atoms undergo diffusive motion characterized by the diffusion coefficient $D$ inversely proportional to $P$, and the diffusion time across the cell is set by $\tau_D = R^2/D$. 
This time is much longer than the time it would have taken to cross the cell in pure ballistic motion, so that the diffusive regime holds for $\tau_D \gg 2R/v$, i.e. $D \ll vR/2$. For helium in our specific cell this corresponds to $P \gg 0.05$~mbar, i.e. the upper part of the parameter space in Fig.~\ref{fig:classification}.

In the diffusive regime, the spectral density of the fluctuating magnetic field seen by the atoms randomly moving in a static gradient can be derived from the diffusion equation. 
Then, following McGregor \cite{McGregor1990}, the Redfield theory of relaxation can be applied to yield expressions for the relaxation rate, which depends on the geometry of the cell and the shape of the magnetic field. 
We derive in detail in the Appendix~\ref{app:cylindrical-cell}  an expression adequate for the case of our study: 
\begin{eqnarray} \label{eq:gamma1-final}
    \Gamma _1= D \frac{G^2}{B^2} \ds\sum _{k=1}^{\infty} \frac{(\gamma B \tau_D)^2}{(\gamma B \tau_D)^2+ x_{1k}^4}\frac{2}{x_{1k}^2-1} ,
\end{eqnarray}
where $\gamma$ is the gyromagnetic ratio ($\gamma = 2\pi \times 7.6~\mathrm{Hz/\upmu T}$ for mercury-199) and $x_{1k}$ is the $k$-th zero of the derivative of the Bessel function $J_1$. 
In fact, Eq. \eqref{eq:gamma1-final} becomes invalid at high Larmor frequency $\gamma B$ because the correlation function is not described by the diffusion equation for times shorter than the time between two atomic collisions $\tau_c = 4D/v^2$. The validity domain of Eq.  \eqref{eq:gamma1-final} is represented by the upper gray zone in Fig.\ref{fig:classification}. 

At moderate field $B$, specifically $\gamma B \tau_D \gg 1$ and $\gamma B \tau_c \ll 1$ (corresponding to the dark dotted zone in Fig.~\ref{fig:classification} labeled \textit{adiabatic diffusive}), the general expression  \eqref{eq:gamma1-final} simplifies to the formula \eqref{eq:gamma1-adiab-diff}. 
Physically, this regime is such that there are many atomic collisions in one Larmor time  $(\gamma B)^{-1}$, but many Larmor times during the time $\tau_D$ it takes to traverse the cell diffusively. 
In the adiabatic diffusive  regime, the relaxation rate becomes independent of the geometry of the cell \cite{Guigue2014d}. 
This is an advantage for determination of the diffusion coefficient, because it reduces systematic errors due to the imperfections of the geometry of the cell.

The rectangular region marked with dashes in Fig.~\ref{fig:classification} is our region of interest, corresponding to the parameters used in the experiment. 
Conceptually, the experiment consists in filling a cell with a mixture of mercury and a gas with a controlled pressure, applying a known field $B$ and a known gradient $G$, measuring the relaxation rate $\Gamma_1$, and deducing an effective diffusion coefficient $D_{\rm eff}$ defined by
\begin{equation} \label{eq:D-eff}
     \Gamma_1(G) \equiv D_{\rm eff} \frac{G^2}{B^2} + \Gamma_1(0).
\end{equation}
It is actually necessary to measure $\Gamma_1$ as a function of the  applied gradient in order to separate the magnetic relaxation from the other sources of relaxation $\Gamma_1(0)$. 
In the region of interest, the effective diffusion coefficient departs from the true coefficient $D$ by up to $30$\%, which is larger than the accuracy of the measurement (about $1\%$, as we will see). 
In fact, based on \eqref{eq:gamma1-final}, we have:
\begin{equation}\label{eq:tilde-D}
    D_{\mathrm{eff}} = \ds\sum _{k=1}^{\infty} \frac{D}{1+\left(\frac{D}{\gamma BR^2}\right) ^2 x_{1k}^4}\frac{2}{x_{1k}^2-1}.
\end{equation}
In practice, as will be explained later, for a given cell $D_{\rm eff}$ was measured for different values of the holding field $B$, then Eq.~\eqref{eq:tilde-D} was used to extract the true diffusion coefficient $D$.

\section{Experimental setup}
\label{sec:experiment}

\begin{figure}
\centering
    \includegraphics[width=0.45\textwidth]{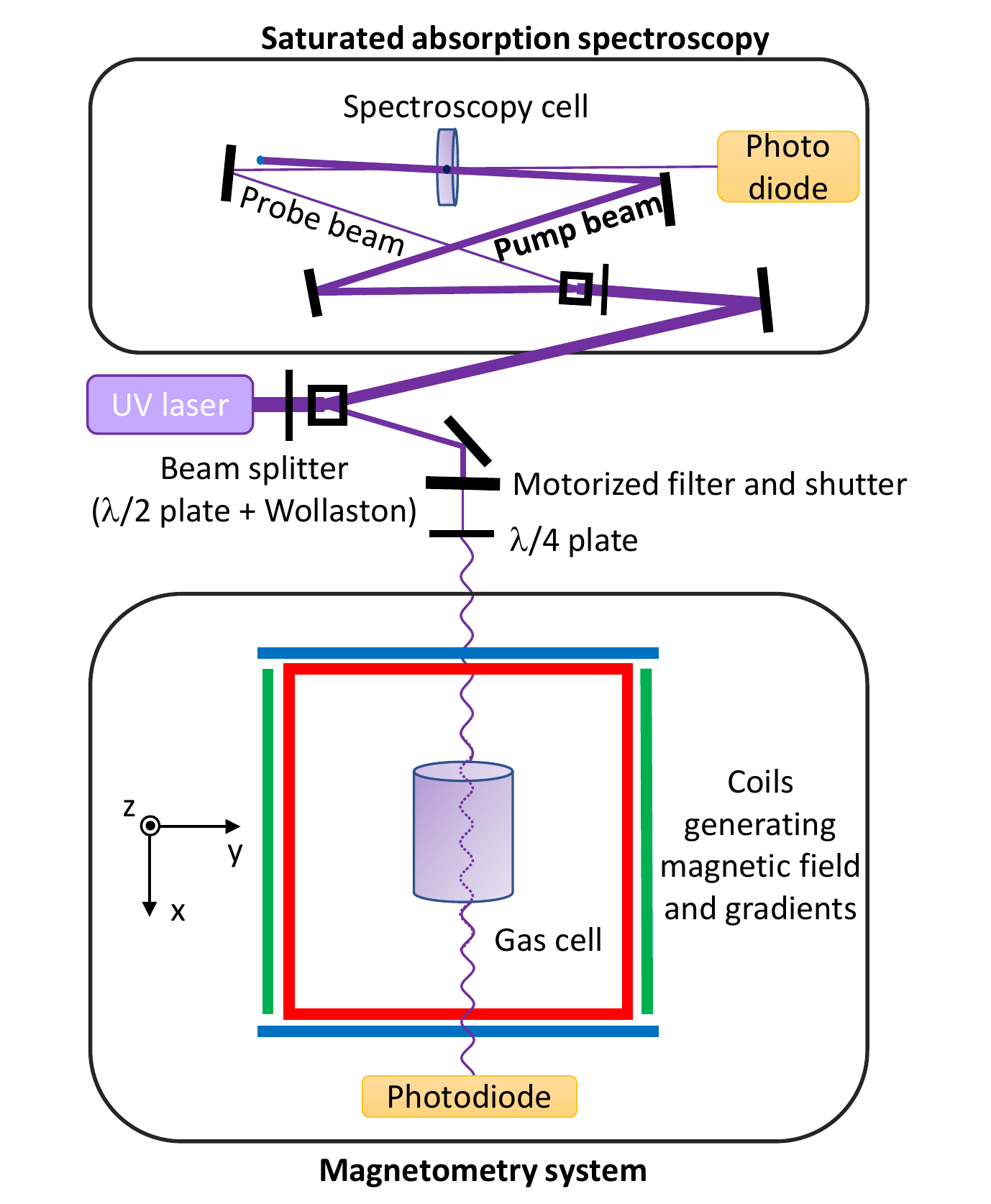}
    \caption{\label{fig:expsetup} A schematic representation of the laser and magnetometry setups. The initial beam is split in two linearly polarized beams. The top one serves for locking the laser frequency on the mercury-199 $F=1/2$ transition obtained through saturated absorption spectroscopy. The bottom one is circularly polarized with a quarter-wave plate and is used for optical pumping and precession measurements.}
\end{figure}

\begin{figure}
\centering
    \includegraphics[width=0.45\textwidth]{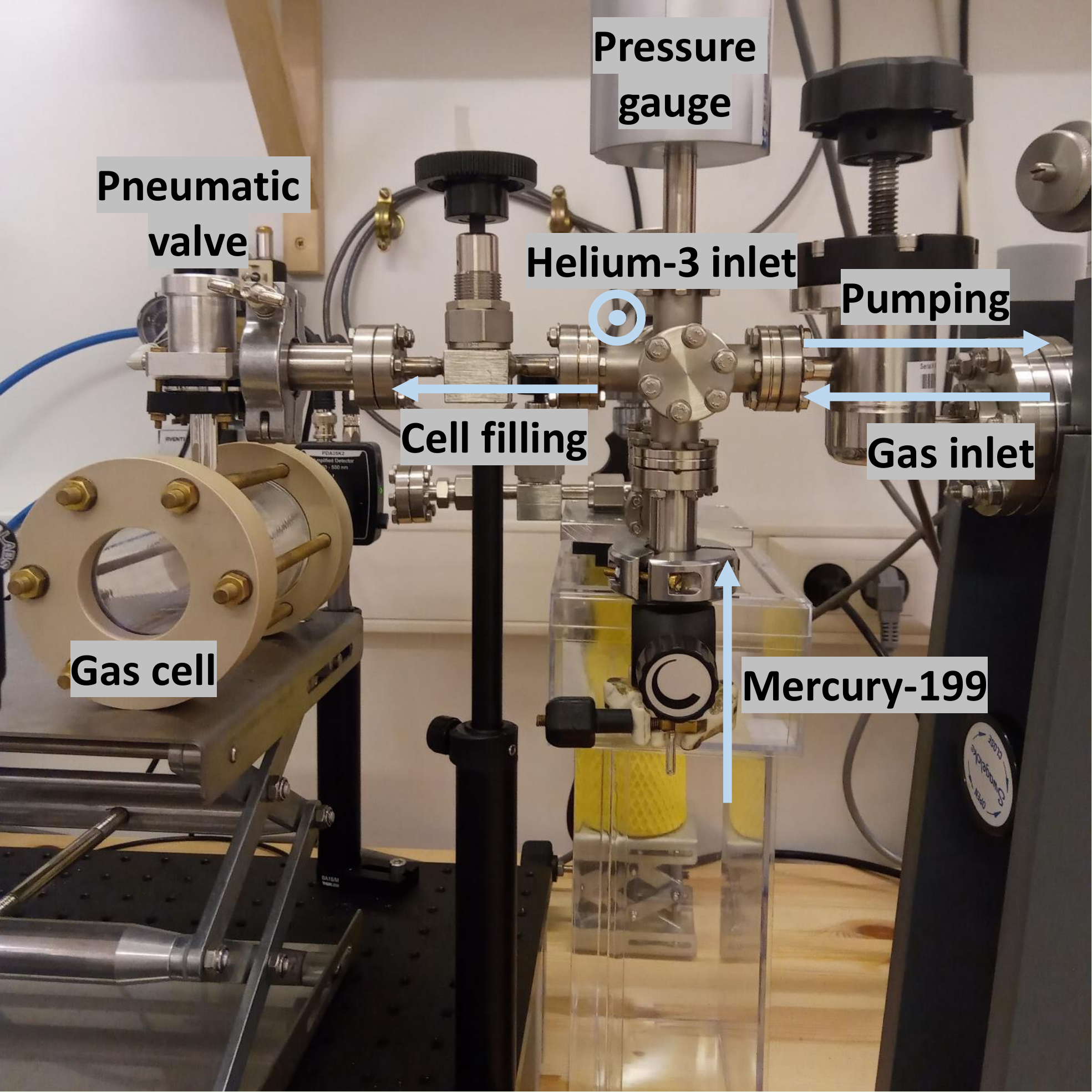}
    \caption{\label{fig:filling} Photograph of the cell filling station, with the magnetometry cell connected. The arrows indicates the various gas flows allowed within the system.}
\end{figure}

The experimental setup is composed of three parts: a 253.7~nm laser, a system generating magnetic fields and gradients which are both schematized in Fig.~\ref{fig:expsetup}, and a cell filling station shown in Fig.~\ref{fig:filling}.

The ground level $^1S_0$ of mercury-199 has a total spin $F=1/2$ arising only from the nuclear spin. The excited $^3P_1$ level at $253.7$~nm is split in two hyperfine sublevels $F=1/2$ and $F=3/2$ separated by $22$~GHz. Mercury atoms can be polarized by optical pumping on the $F=1/2$ to $F=1/2$ transition. A $253.7$~nm TOPTICA TA-FHG pro laser is used to realize the optical pumping. It consists in an infrared laser diode coupled to two frequency doubling non linear crystals. A grating allows for a fine variation of the laser frequency around the resonance. Saturated absorption spectroscopy provides a reference to lock the laser on the precise frequency of the $F=1/2$ to $F=1/2$ transition of mercury-199.
The laser output power is at maximum 20~mW. The beam is split in two, most of the power is used for the laser locking scheme, and only 30 to $60~\mathrm{\upmu W}$  are used for the experiment itself.
Using half-wave and quarter-wave plates, the laser beam is polarized circularly for the experiment. A motorized filter and a shutter allow for fast (a few tens of milliseconds) variation of the beam intensity.
The  polarized beam is directed through a spectroscopy cell filled with mercury and a few millibar of buffer gas. 

Our experimental cell is placed at the center of three pairs of orthogonal square coils.
Each pair of coils is capable of generating a magnetic field up to $200~\muT$. The coils are used to cancel the ambient magnetic field, set the fields for optical pumping and free precession, and generate magnetic gradients. 
We generate the magnetic gradient by controlling separately the current in the two coils forming the ``$x$" pair. 
In the volume of the cell the magnetic field takes the form:
\begin{equation}\label{eq:field-shape}
\vec{B}(x,y,z) = \begin{pmatrix}
B_x(0) \\
B_y(0) \\
B_z(0) \\
\end{pmatrix} + G_x \begin{pmatrix}
x \\
-y/2 \\
-z/2 \\
\end{pmatrix}. 
\end{equation}
When the field is set along the $x$ axis, the depolarizing gradient in  \eqref{eq:gamma1-final} is $\displaystyle G = \sqrt{\left(\vec\nabla B_y\right)^2 + \left(\vec\nabla B_z\right)^2} = \frac{G_x}{\sqrt{2}}$. 
We performed a magnetic mapping in the central region (without the cell) to calibrate the magnetic generation system before the experiment, and repeated the mapping after the experiment. 
The uncertainty on the set field is $\Delta B_x = 0.5~\muT$ and the relative uncertainty on the set gradient is $\Delta G_x/G_x = 1~\%$. 
As the coils power supplies are uni-polar, the maximum gradient that can be applied  is imposed by the earth's magnetic field, and goes from $G_x=\pm300~\mathrm{nT/cm}$ at $B_x=15~\muT$ to $G_x=\pm1400~\mathrm{nT/cm}$ at $B_x=120~\muT$.
After crossing the cell, the light is detected using a photodiode whose signal is digitized through a 16 bit 
ADC. NI Labview is used to integrate and coordinate the control of the various subsystems: field and gradient generation, UV light intensity and data recording.

The cell used in the experiment is a 10~cm long glass cylinder with an internal diameter of 4~cm. It is closed on both ends with quartz windows.
A small glass tube ended with a pressure-controlled valve allows to connect the cell to a filling station through a KF16 flange.
The filling station consists in a small volume to which several items isolated by valves are connected: a turbo-pump to empty the whole system, a pressure gauge, a small volume containing a droplet of enriched mercury-199 and two gas inlets. The valve for these last devices allows for a fine control of the opening, allowing to adjust precisely the pressure of gas within the system.
One inlet is connected to a small helium-3 bottle that is never removed whereas the gas bottle on a second inlet can be changed.
The pressure gauge is a capacitance manometer that gives absolute measurements with a microbar accuracy in the microbar to tens of millibar range, independently of the nature of the gas. 
To prepare a cell, the full system is first evacuated, then filled with the desired gas, close to the targeted pressure. The valve to the mercury droplet is then opened and we wait for mercury at saturated vapor pressure (at ambient temperature) to diffuse into the cell. The quantity of mercury is controlled through the absorption of UV light. When the absorption is within $45\pm5\%$, the cell is isolated and disconnected from the filling station.
Either the pneumatic valve or the toric seals at the cell windows have small leaks. These leaks are monitored by measuring the pressure within the cell at the end of each measurement. 

\section{Analysis procedure}
\label{sec:analysis}
\paragraph*{Data taking strategy.}
\label{sec:data-taking-strategy}

\begin{figure}
\centering
    \includegraphics[width=0.45\textwidth]{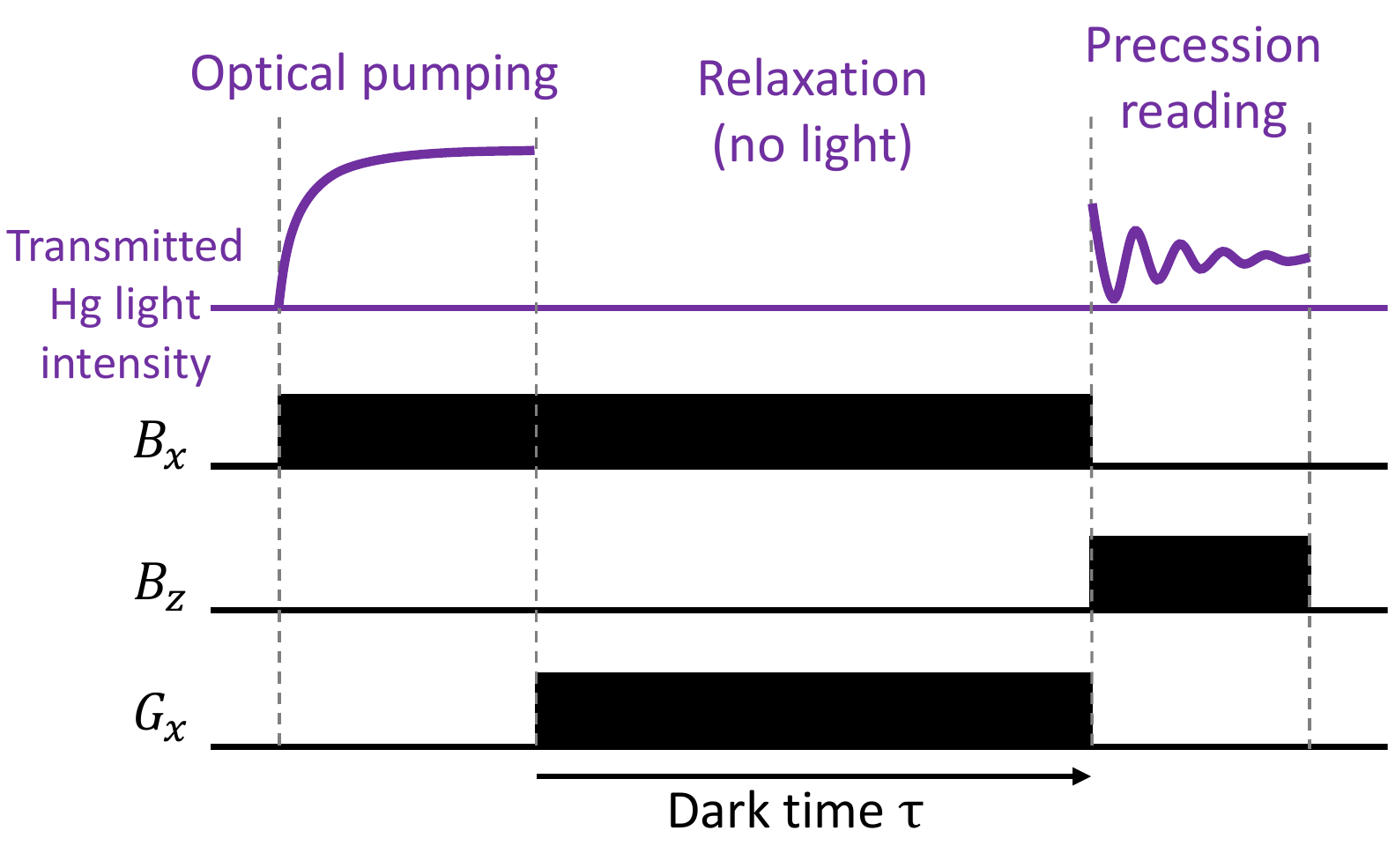}
    \caption{\label{fig:timtable}General time structure of the polarization measurement sequence for the transmitted laser intensity, the magnetic field and the gradient. The transitions between the main steps are not indicated.}
\end{figure}

The basic element of the data taking is a sequence for a set of parameters $(\tau,B_x,G_x)$ consisting in:
\begin{itemize}
\item Optical pumping at $60~\mathrm{\upmu W}$, with a holding magnetic field along the $x$-axis (of magnitude $B_x$) for 8~s,
\item Closure of the beam shutter and setting of a magnetic gradient $G_x$ along the $x$-axis over $0.2$~s,
\item Relaxation during a time $\tau$ in the dark. This is during this time that the depolarization we want to measure occurs,
\item Opening of the beam shutter with the gradient at zero and setting of a field along the $z$-axis (of magnitude $B_z = 40~\muT$). We take care that during the transition, the magnetic field never goes to zero to avoid depolarization. The change of direction over $0.4$~s of the magnetic field is fast enough so the polarization stays along the $x$-axis and can precess in the $xy$ plane,
\item Recording of the precession  signal in the $40~\muT$ field for 3~s, with a light power reduced to $30~\mathrm{\upmu W}$,
\item Setting of the field to zero to depolarize the mercury atoms.
\end{itemize}
The time structure of a sequence is illustrated on Fig.~\ref{fig:timtable}.
The analysis of each sequence will give a value of the polarization after a dark time $\tau$. 
Five consecutive sequences make a $\tau$-scan for a set $(B_x,G_x)$ with varying dark time.
From a $\tau$-scan one extracts a value of $\Gamma_1$.
Using short $\tau$-scans of 2 to 2.5~min makes the analysis insensitive to drifts of the laser intensity as well as variations of the cell absorption which have a larger timescale. 
A set of seven consecutive $\tau$-scans for a fixed $B_x$, varying $G_x$, is a gradient scan from which we extract a value of $D_{\mathrm{eff}}$. 
A gradient scan lasts for approximately 15~min making the analysis insensitive to slow drifts of the ambient field.
A full run is a collection of gradient scans performed several times for each value of $B_x$.

\paragraph*{Precession signals and extraction of the polarization from a sequence.}

In order to extract the longitudinal relaxation rate $\Gamma_1$ from a $\tau$-scan, the polarization $p$ of the atoms at the beginning of the precession must be measured from sequences with different values of dark times $\tau$, as shown in Fig.~\ref{fig:timtable}. 
The precession signal, that is, the power of the transmitted light, is not a simple damped-oscillator signal, but is given by:
\begin{equation}
    S(t) = S_0 \exp \left( -n_{\mathrm{Hg}}\sigma _{\hg} (1+p_{\parallel}(t)) L \right),
\end{equation}
where $S_0$ is the light power  entering the cell, $n_{\hg}$ the Hg atoms density in the cell, $\sigma _{\hg}$ the light absorption cross-section and $L$ the optical path length, and $p_{\parallel}(t)$ is the instantaneous projection along the beam of the precessing atomic polarization: \begin{equation}
    p_{\parallel}(t) = p \cos (\gamma B_z t)  e^{-\Gamma_2 t}.
\end{equation}
Here $\Gamma_2$ accounts for the decay of the transverse polarization during the precession. 
Then, the polarization $p$ is obtained from the ratio of the maximum $S_{\rm max}$ to minimum $S_{\rm min}$ of the signal $S$ by:
\begin{equation}\label{eq:extracted-polarization}
    p = \frac{\ln \frac{S_{\mathrm{max}}}{S_{\mathrm{min}}}}{-\ln (1-A)}.
\end{equation}
In the expression \eqref{eq:extracted-polarization}, $A = 1 - \exp \left(-n_{\hg}\sigma _{\hg}L\right)$ is the total absorption of the cell with unpolarized atoms, which is  measured at the beginning (and also at the end) of the run when filling the cell. 

For each run, we provide a file in the supplementary material with the results of the analysis of each sequence, including in particular $S_\mathrm{min}$ and $S_\mathrm{max}$.


\paragraph*{Extraction of the longitudinal relaxation rate from a $\tau$-scan.}

\begin{figure}
    \includegraphics[width=0.45\textwidth]{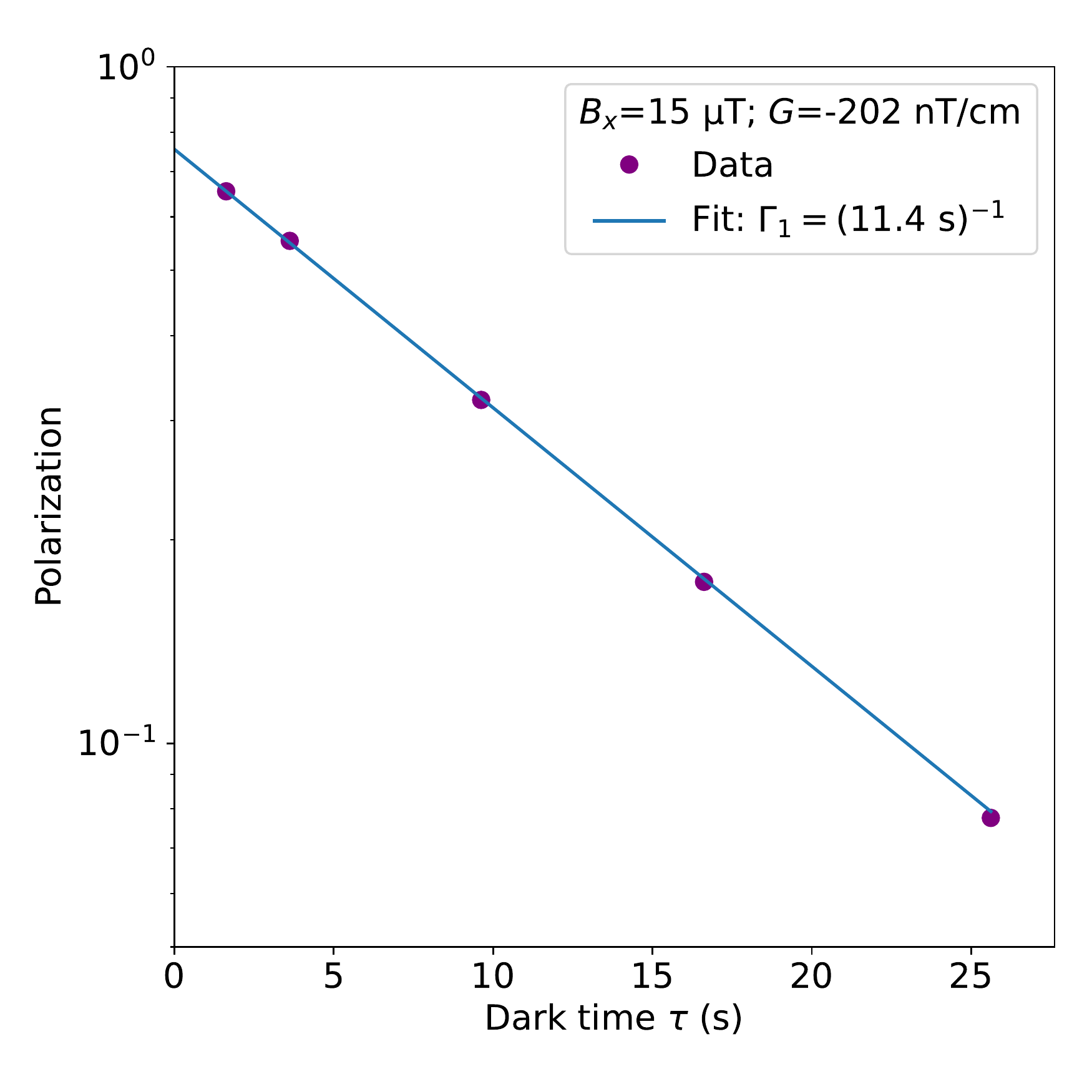}
    \caption{\label{fig:exp-T1}Example of polarization evolution with the dark time for a holding magnetic field of $15~\muT$ and a gradient of $-202~\mathrm{nT/cm}$.
    The cell is filled with $1~\mbar$ of helium-4.
    The time constant $\Gamma_1$ of the exponential model is obtained by fitting the data.}
\end{figure}

The polarization $p$ is measured for several dark times $\tau$ in order to produce a $\tau$-scan.
Figure~\ref{fig:exp-T1} shows an example of $\tau$-scan for a given configuration of magnetic field for a cell of helium-4 at $1~\mbar$.
An exponential model is then fitted to extract the relaxation rate $\Gamma_1$ for each $\tau$-scan.


\paragraph*{Extraction of $D_{\mathrm{eff}}$ from a gradient scan.}

\begin{figure}
    \includegraphics[width=0.45\textwidth]{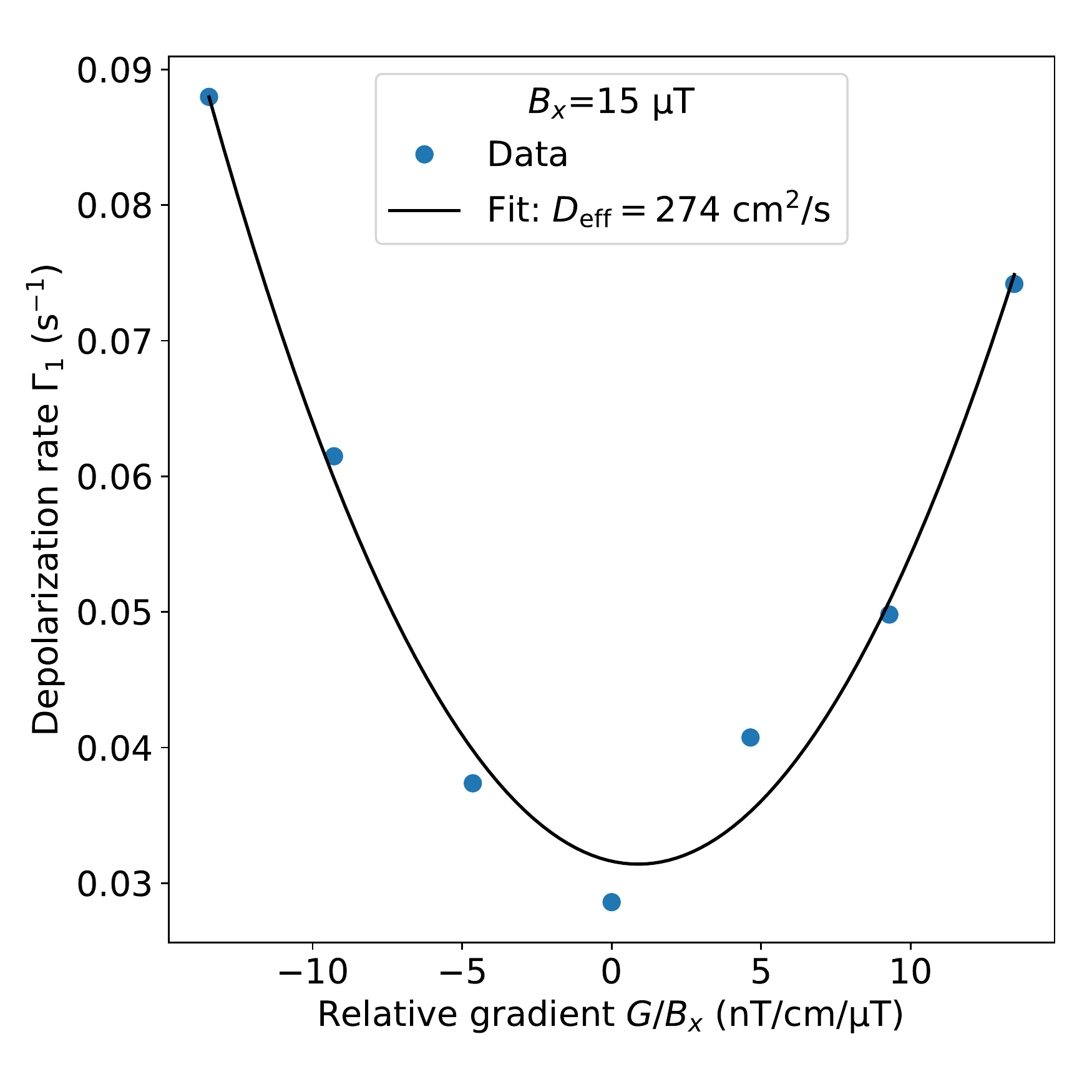}
    \caption{\label{fig:one-parabola}Longitudinal relaxation rate $\Gamma_1$ extracted from a gradient scan for a cell of $1~\mbar$ of helium-4 as a function of the relative transverse gradient $G/B_x$.
    Data are fitted using Eq.~\eqref{eq:parabola} to extract $D_{\mathrm{eff}}$.}
\end{figure}

The extraction of the longitudinal relaxation rate is repeated for several values of transverse magnetic gradients $G$ while keeping the same value of the holding magnetic field $B_x$.
Figure \ref{fig:one-parabola} shows the relaxation rate $\Gamma_1$ obtained for seven values of magnetic field gradient between $-290~\mathrm{nT/cm}$ and $290~\mathrm{nT/cm}$.

This collection of measurements represents a parabola whose curvature corresponds to the diffusion coefficient $D_{\mathrm{eff}}$ described in Eq.~\eqref{eq:D-eff}. 
In fact we fit with a general parabola using~: 
\begin{equation}\label{eq:parabola}
    \Gamma_1(G) = \Gamma_1(0)+C\times \frac{G}{B_x} + D_{\mathrm{eff}}\times \frac{G^2}{B_x^2}.
\end{equation}
where  $\Gamma_1(0)$, $C$ and $D_{\mathrm{eff}}$ are the adjusted parameters.
The coefficient $\Gamma_1(0)$ corresponds to all the field-independent depolarization processes, such as the collision of the \hg with the cell walls or with other atoms in the gas.
It also includes the small magnetic relaxation rate contribution induced by magnetic gradients produced by and proportional to the holding magnetic field. 
The linear term $C$ is the cross-term that accounts for a possible offset of the magnetic gradient, after the analysis this term is found to be compatible with zero. 
The average values of $1/\Gamma_1(0)$ for each run are listed in Tab.~\ref{tab:liste-D}. 
The values are all similar and of the order of 40~s except for $\mathrm{O}_2$ and air, due to the paramagnetic nature of the oxygen molecule.

\begin{table}[b]
\centering
\caption{Gradient-independent relaxation time $1/\Gamma_1(G=0)$ and diffusion coefficient $D$ extracted from each run.
The correction applied to each value (right column) is detailed in Sec.~\ref{sec:extraction-D}.
    }
	\label{tab:liste-D}
    \begin{tabular}{c c c c c}
        \hline\hline
        Gas                 & Initial pressure  & $1/\Gamma_1(G=0)$       & $D$                 & Corrected $D$       \\
                            & (mbar)            & ($\mathrm{s}$)   & (\cmsps)            & (\cmsps)            \\ \hline
        $^3\mathrm{He}$     & 1.0               & 39.7(3)          & 578(10)             & 617(12)             \\
        $^3\mathrm{He}$     & 2.0               & 40.3(3)          & 267(5)              & 289(6)              \\
        $^3\mathrm{He}$     & 5.0               & 43.5(4)          & 119(6)              & 121(6)              \\ \hline
        $^4\mathrm{He}$     & 1.0               & 34.6(4)          & 495(9)              & 519(10)              \\
        $^4\mathrm{He}$     & 2.0               & 36.4(1)          & 262(5)              & 265(5)              \\
        $^4\mathrm{He}$     & 5.0               & 36.6(1)          & 105(4)              & 107(4)              \\ \hline
        $\mathrm{Ar}$       & 1.0               & 39.2(5)          & 110(7)              & 111(7)              \\
        $\mathrm{Ar}$       & 1.0               & 38.5(4)          & 111(7)              & 112(7)              \\ \hline
        $\mathrm{Kr}$       & 0.5               & 36.1(1)          & 149(8)              & 150(8)              \\
        $\mathrm{Kr}$       & 1.0               & 36.4(3)          & 75(5)               & 75(5)               \\ \hline
        $\mathrm{Xe}$       & 0.5               & 39.4(6)          & 106(5)              & 107(5)              \\
        $\mathrm{Xe}$       & 1.0               & 41.2(3)          & 41(5)               & 41(4)               \\ \hline
        $\mathrm{N_2}$      & 0.5               & 34.2(4)          & 257(7)              & 265(7)              \\
        $\mathrm{N_2}$      & 1.0               & 37.7(4)          & 115(6)              & 116(6)              \\
        $\mathrm{N_2}$      & 2.0               & 35.8(4)          & 71(4)               & 73(4)               \\
        $\mathrm{N_2}$      & 2.0               & 43.9(6)          & 57(5)               & 57(5)               \\ \hline
        $\mathrm{CO_2}$     & 0.5               & 44.2(4)          & 174(5)              & 178(5)              \\
        $\mathrm{CO_2}$     & 1.0               & 42.6(4)          & 90(6)               & 91(6)               \\ \hline
        $\mathrm{O_2}$      & 0.2               & 7.22(8)          & 577(35)             & 598(37)             \\
        $\mathrm{O_2}$      & 0.3               & 4.63(4)          & 339(76)             & 340(77)             \\ \hline
        $\mathrm{Air}$      & 0.5               & 9.50(4)          & 235(10)             & 243(11)             \\
        $\mathrm{Air}$      & 1.0               & 5.40(5)          & 88(28)              & 89(29)              \\
        $\mathrm{Air}$      & 1.0               & 5.73(5)          & 177(18)             & 180(19)             \\ \hline
    \end{tabular}
\end{table}

\paragraph*{Extraction of $D$ from a run.}
\label{sec:extraction-D}

\begin{figure}
    \includegraphics[width=0.45\textwidth]{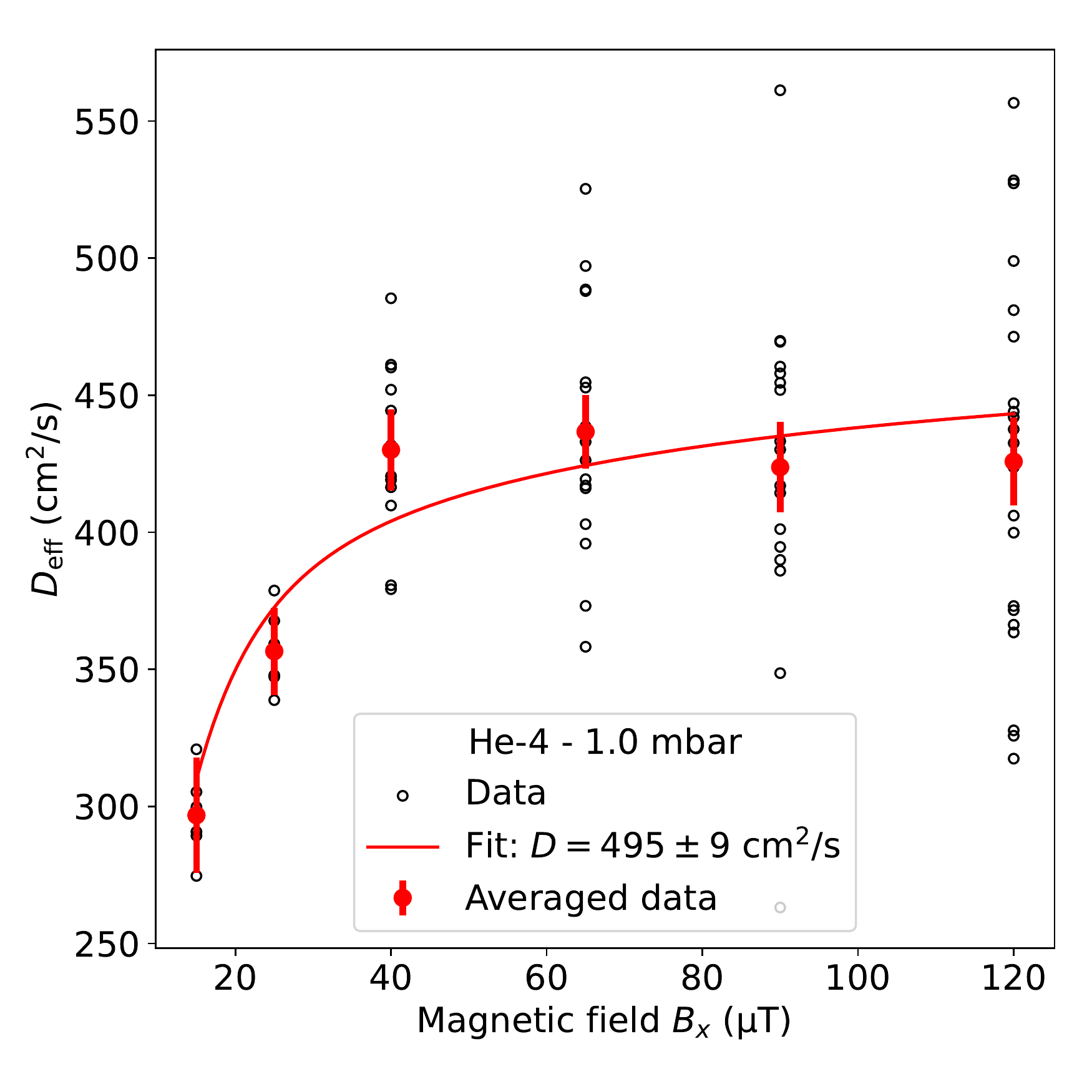}
    \caption{\label{fig:moneyplot}Parameter $D_{\mathrm{eff}}$ obtained from several gradient scans as a function of the holding magnetic field $B_x$ for a $1~\mbar$ helium-4 cell (black circles).
    The red dots represents the average value and error for each $B_x$.
    The model \eqref{eq:tilde-D} (red line) is fitted to the data in order to extract the diffusion coefficient $D$.}
\end{figure}

As shown in Fig. \ref{fig:moneyplot}, for a given run, we get several values of $D_{\rm eff}$ for each value of the holding magnetic field $B_x$ (black circles), that we average to get the red dots. 
The previous fits are performed using a linear regression, all the points having the same weights and no errors are computed. For a given holding field, the error on the average $D_{\rm eff}$ in fig.\ref{fig:moneyplot} is computed from the standard deviation of $D_{\rm eff}$ divided by the square root of the numbers of measurement. This error captures the statistical precision as well as the imperfections of the fitting procedure.

The number of repetitions for each holding field value $B_x$ was optimized in order to have a statistical error almost identical for each point. 
The statistical fluctuation on the parameter $D_{\mathrm{eff}}$ increases with $B_x$ since the possible range of $G/B_x$ gets smaller with larger $B_x$.
In addition, the magnetic field and gradient are not perfectly known, which leads to systematic errors on the extracted $D_{\mathrm{eff}}$. 
The error bars represented on the averaged values of $D_{\mathrm{eff}}$ (red dots) include both statistical and systematic errors in quadratic sum.

Finally, we can extract the value and error of the parameter $D$ from Eq.~\ref{eq:tilde-D} by fitting this model to the data of each run using a $\chi ^2$ minimization method.
Tab. \ref{tab:liste-D} presents the extracted diffusion coefficient $D$ for each run. 
Figures analogous to fig. \ref{fig:moneyplot} are provided in the supplementary material for all the runs. 

The cell is known to leak with the pressure increasing linearly over time at an approximate rate of $0.8~\mathrm{\upmu bar/h}$.
The average quantity of air in the cell over the measurement is determined by recording the pressure within the cell before and after the measurement.
We consider a stagnant mixture of mercury in some gas at partial pressure $P_\mathrm{gas}$ with an additional quantity of air (treated as a single species) at partial pressure $P_{\mathrm{air}}$. The concentration of mercury is negligible with respect to the other gases.
Then the effective diffusion coefficient of mercury in the mixture is~\cite{Hirschfelder1955}:
\begin{equation}
\frac{P_\mathrm{mix}}{D^{\mathrm{mix}}(P_\mathrm{mix}) }\approx \frac{P_{\mathrm{air}}}{D^{\mathrm{air}}(P_\mathrm{mix})}+\frac{P_\mathrm{gas}}{D^\mathrm{gas}(P_\mathrm{mix})}
\end{equation}
where $P_\mathrm{mix} = P_{\mathrm{air}}+P_\mathrm{gas}$ is the total pressure of the mixture. As $D(P)P$ is a constant, inverting this formula relates the diffusion coefficient of mercury in the gas $D^\mathrm{gas}(P_\mathrm{gas})$ to the measured $D^{\mathrm{mix}}(P_\mathrm{mix})$:
\begin{equation}\label{eq:correction-D}
\frac{1}{D^\mathrm{gas}(P_\mathrm{gas})} = 
\frac{1}{D^{\mathrm{mix}}(P_\mathrm{mix})}-\frac{P_{\mathrm{air}}}{D^{\mathrm{air}}(P_0)P_0}
\end{equation}
where the diffusion coefficient $D^{\mathrm{air}}(P_0)$ at some reference pressure $P_0$ is measured independently. 
The diffusion coefficients corrected using Eq.~\eqref{eq:correction-D} are reported in the last column of Tab.~\ref{tab:liste-D}.
We use the combination of the measurements of diffusion coefficient in air listed in Tab.~\ref{tab:final-results} to implement this correction. The error on this measurement is propagated on all the final results.
As there are three independent measurements for air, all three are slightly modified by this correction but the overall combination is not. 

\begin{table}[b]
    \centering
    \caption{Comparison between the diffusion coefficient $D$ extracted from our measurements, theory (Chapman-Enskog) and other results from the literature.
    The values displayed with an asterisk \cite{Massman1999} are a combination of other measurements.
    The values expressed in \cmsps  are extrapolated at 1~mbar pressure and 293~K.}
        \label{tab:final-results}
        \begin{tabular}{c c c c}
            \hline\hline
            Gas                 & This work     & Theory   & Literature                             \\ \hline
            $^3\mathrm{He}$     & 598(8)        & 786   &                                               \\
            $^4\mathrm{He}$     & 523(7)        & 693   & 566 \cite{Gardner1991}                        \\
            $\mathrm{Ar}$       & 111(5)        & 113   & 113 \cite{Gardner1991}, 147 \cite{Spencer1969} \\
            $\mathrm{Kr}$       & 75(3)         & 70    & 69(4) \cite{Nakayama1968}                     \\
            $\mathrm{Xe}$       & 64(4)         & 50    & 49(3) \cite{Nakayama1968}                     \\
            $\mathrm{N_2}$      & 129(3)        & 129   & 145 \cite{Gardner1991}, 121(9) \cite{Nakayama1968}, 138$^*$ \cite{Massman1999} \\
            $\mathrm{CO_2}$     & 89(2)         & 80    & 113 \cite{Gardner1991}                        \\
            $\mathrm{O_2}$      & 118(7)        & 123   &                                               \\
            $\mathrm{Air}$      & 125(5)        & 128   & 136$^*$ \cite{Massman1999}                    \\   
           $\mathrm{Air}$\footnote{This value is obtained by combining our measurements of nitrogen, oxygen, argon, and  carbon dioxide.}     
                                &  126(3)        & -    & -                    
                \\
            \hline
        \end{tabular}
    \end{table}

\section{Results and discussion}
\label{sec:results}

For each gas, several runs have been performed for different gas pressure as summarized in Tab.~\ref{tab:liste-D}. The dependency as an inverse of the pressure is verified and we combine all available data for each gas to compute the diffusion coefficient for a pressure of $1~\mbar$. These are reported in Tab.~\ref{tab:final-results}.

A theoretical model for binary diffusion coefficients can be obtained from the kinetic theory of gases. 
The formula derived independently by Chapman and Enskog~\cite{Pauling2001, Hirschfelder1955, Chapman1939} relies on a Lennard-Jones interaction with parameters $\sigma = \sigma_1+\sigma_2$ and $\varepsilon=\sqrt{\varepsilon_1\varepsilon_2}$, where $\sigma_1$, $\sigma_2$, and $\varepsilon_1$, $\varepsilon_2$ are respectively the diameter and the depth of the individual  Lennard-Jones potential for each species.
The Chapman-Enskog formula is then:
\begin{equation}
    D=\frac{3kT}{8P\pi\sigma^2\Omega^*(\varepsilon/T)}\left(\frac{\pi kT (m_1+m_2)}{2m_1m_2}\right)^\frac{1}{2}
\end{equation}
where $m_1$ and $m_2$ are the masses of the two species, $T$ the temperature, and $\Omega^*$ the collision integral, quantifying the correction from the hard sphere model, which depends on the depth of the Lennard-Jones potential and the temperature. To compute the values given in Tab.~\ref{tab:final-results}, we use the Lennard-Jones parameters derived from viscosity data listed in~\cite{Pauling2001} and the $\Omega^*$ tabulated in~\cite{Hirschfelder1955}. 

Our measurements are in a rather  good agreement with the theoretical calculations, except for helium. 
This is to be expected as Lennard-Jones potential does not model correctly the behavior of mercury helium mixture~\cite{Mueller1963}. Nevertheless, the ratio of diffusion coefficient between helium-3 and helium-4 should only depend on the masses:
\begin{equation}
r^2 = \left(\frac{D^{^3\he}}{D^{^4\he}}\right)^2\approx\frac{m_{^4\he}}{m_{^3\he}}\approx1.33.
\end{equation}
Indeed our measurement leads to $r^2=1.31\pm0.05$ which is in agreement with the expected value.

We also report in Tab.~\ref{tab:final-results} the results of other measurements (using non-NMR methods) available in the literature. 
The values are extrapolated to the temperature of $293$~K using the Chapman-Enskog temperature dependence $\frac{1}{\Omega^*}T^{\frac{3}{2}}$. 
Although the measurements are not all compatible within the quoted errors, our set is in general in rather good agreement with the previous determinations. 

In conclusion, we have developed a method to measure the diffusion coefficient of mercury in a generic gas at room temperature and low pressure, which led to a consistent set of measurements for nine gases, including helium-3 and oxygen which were not measured before. 

\section*{Acknowledgments}
The authors would like to thank F.~Lalo\"e, P.~J.~ Nacher and G.~Tastevin for helpful discussions and genuine interest. 
Many of the gas bottles used in the experiment were kindly provided by A.~Bes, O.~Guillaudin, M.~Heusch, and A. Palacios-Laloy. 
We acknowledge financial support from the ERC under Project No. 716651-NEDM and from the Labex Enigmass under Project Goldorak.

\appendix

\section{Computation of the longitudinal relaxation rate for a cylindrical cell in uniform magnetic gradients}
\label{app:cylindrical-cell}

In the motional narrowing regime of spin relaxation, the effect of a random perturbation can be evaluated with the help of second order perturbation theory \cite{Abragam1961}. 
The longitudinal depolarization rate $\Gamma_1$ can be written as:
\begin{equation}\label{eq:Gamma1-def}
    \Gamma_{1}=\gamma^{2} \int_{0}^{\infty} \operatorname{Re}\left[\left\langle b^{*}(0) b(t)\right\rangle e^{i \omega t}\right] \mathrm{d} t, 
\end{equation}
where $\omega$ is the Larmor angular precession frequency and $b$ is a complex noise representing the transverse components of the magnetic field $\vec{B}$. 
If the main magnetic field is taken to be along $x$, then $b=B_y + iB_z$. 

Now, we have to compute the correlation function of $b$. 
Following the work of McGregor \cite{McGregor1990}, we write:
\begin{equation} \label{eq:correlation-b}
\left\langle{b}^{*}(0) {b}(t)\right\rangle=\int_V \frac{1}{V} \mathrm{d} \overrightarrow{r_{0}} \int_V d \vec{r} b^{*}\left(\overrightarrow{r_{0}}\right) b(\vec{r}) \pi\left(\vec{r}, t \mid \overrightarrow{r_{0}}\right),
\end{equation}
where $V$ is the volume of the cell and $\pi(\vec{r},t\mid\vec{r}_{0})$ is the conditional probability (or propagator) for an atom  to be at $\vec{r}$ at the time $t$ given that it started at the position $\vec{r}_{0}$ at $t=0$.

In the case of the diffusive regime, it is possible to obtain an analytical expression for the propagator. 
Indeed the propagator follows the diffusion equation:
\begin{equation}
    D \Delta \pi\left(\vec{r}, t \mid \overrightarrow{r_{0}}\right)=\frac{\partial}{\partial t} \pi\left(\vec{r}, t \mid \overrightarrow{r_{0}}\right),
\end{equation}
where $D$ is the diffusion coefficient.
Then, $\pi(\vec{r},t\mid\vec{r}_{0})$  must satisfy the initial condition:
\begin{equation}
    \pi\left(\vec{r}, t=0 \mid \overrightarrow{r_{0}}\right)=\delta\left(\vec{r}-\overrightarrow{r_{0}}\right)\label{eq:initial-condition},
\end{equation}
and the boundary condition on the cell walls:
\begin{equation}\label{eq:boundary-condition}
    \vec{\nabla} \pi\left(\vec{r}, t \mid \overrightarrow{r_{0}}\right) \cdot \vec{n}=0
\end{equation}
where $\vec{n}$ is the vector normal to the surface.

Let us consider the diffusion within a cylindrical cell of length $L$ and radius $R$ where its axis is aligned with the $x$ axis.
A convenient coordinate system is the cylindrical one $(\rho, \theta, x)$: let us use the method of separation of variables and search for solutions in the form of:
\begin{eqnarray*}
    f(\rho, \theta, x,t) = \mathcal{R} (\rho) \mathcal{A} (\theta) \mathcal{X}(x) \mathcal{T}(t).
\end{eqnarray*}
When injecting this equation into the diffusion equation, we obtain:
\begin{eqnarray*}
    \mathcal{R}(\rho) &=& \mathcal{R}_0J_n(k\rho),\\
    \mathcal{A} (\theta) &=& \mathcal{A}_0 e^{im\theta}, \\
    \mathcal{X}(x) &=& \mathcal{X}_1 \cos(lz) + \mathcal{X}_2\sin(lz),\\
    \mathcal{T}(t) &=& \mathcal{T}_0 \exp \left(-D(k^2+l^2)t\right).
\end{eqnarray*}
where $J_n$ is the cylindrical Bessel function of order $n$.

The boundary conditions \eqref{eq:boundary-condition} on the cell imposes that $k$ and $n$ should be integers and:
\begin{eqnarray}
    \mathcal{R}(\rho) &=& \mathcal{R}_0J_n\left(x_{nk}\frac{\rho}{R}\right),\\
    \mathcal{X}(x) &=& \mathcal{X}_1 \cos\left(n\pi\frac{z}{L}\right) + \mathcal{X}_2\sin\left(n\pi\frac{z}{L}\right),
\end{eqnarray}
where $x_{nk}$ is the $k$-th zero of $J_n$ derivative.

The general solution of the diffusion equation in the cylindrical cell is then the linear combination of all the solutions found above, whose coefficients can be computed using the initial condition \eqref{eq:initial-condition}.
In the end we obtain the conditional probability $\pi\left(\vec{r}, t \mid \overrightarrow{r_{0}}\right)$:
\begin{equation}\label{eq:calcul-propagator}
\pi\left(\vec{r}, t \mid \overrightarrow{r_{0}}\right)=\pi_{\perp}\left(\rho, \theta, t \mid \rho_{0}, \theta_{0}\right) \pi_{\|}\left(x, t \mid x_{0}\right),
\end{equation}
with 
\begin{widetext}
\begin{eqnarray*}
& &\pi_{\perp}\left(\rho, \theta, t \mid \rho_{0}, \theta_{0}\right)= \frac{1}{2\pi R^2}\sum _{m\in \mathbb{Z}, k\in\mathbb{N}^*} \Lambda_{mk,1}^{-1}J_m\left( x_{mk}\frac{\rho }{R}\right)J_m\left( x_{mk}\frac{\rho _0}{R}\right) e^{im(\theta -\theta _0)} e^{-D\left( \frac{x_{mk}}{R}\right) ^2 \vert t\vert},\\
& &\pi_{\|}\left(x, t \mid x_{0}\right)=  \frac{1}{L}+\frac{2}{L}\sum _{n=2,4,...} \cos \left( n\pi\frac{ x}{L}\right)\cos \left(n\pi\frac{ x_0}{L}\right) e^{ \left(-D \frac{n\pi}{L}\right) ^2 \vert t\vert}+\frac{2}{L}\sum _{n=1,3,...} \sin \left( n\pi\frac{x}{L}\right)\sin \left(n\pi\frac{x_0}{L}\right)e^{ \left(-D \frac{n\pi}{L}\right) ^2\vert  t\vert},
\end{eqnarray*}
with $\displaystyle \Lambda_{mk,\alpha}=\frac{1}{R^{\alpha+1}}\left[\int_{0}^{R} J_{m}\left(x_{mk}\frac{ \rho}{R}\right) \rho^{\alpha} d \rho\right]$.
\end{widetext}

Let us now return to the expression \eqref{eq:correlation-b}. 
In the case were $b$ is a linear function of the spatial coordinates, we have:
\begin{equation}\label{eq:simple-gradient}
    b(\vec{r})=\frac{\partial b}{\partial x} x+\frac{\partial b}{\partial y} y+\frac{\partial b}{\partial z} z.
\end{equation}
By injecting Eq.~\eqref{eq:simple-gradient} in Eq.~\eqref{eq:correlation-b} and using Eq.~\eqref{eq:calcul-propagator}, we get:
\begin{equation}\label{eq:computed-correlation-b}
\left\langle{b}^{*}(0) {b}(t)\right\rangle=\left|\frac{\partial b}{\partial x}\right|^{2}\langle{x}(0) {x}(t)\rangle+\left(\left|\frac{\partial b}{\partial y}\right|^{2}+\left|\frac{\partial b}{\partial z}\right|^{2}\right)\langle{z}(0) {z}(t)\rangle
\end{equation}
where
\begin{eqnarray*}
    \langle x(0)x(t )\rangle &=& \frac{8L^2}{\pi ^4}\sum _{n=1,3,...} \frac{1}{n^4}e^{-D\left(\frac{n\pi}{L} \right) ^2\vert t\vert}, \\
    \langle z(0)z(t )\rangle &=&  R^2\sum _{k\in\mathbb{N}^*} \frac{\Lambda_{1k,2}}{\Lambda_{1k,1}} e^{-D\left(\frac{x_{1k}^2}{R} \right) ^2\vert t\vert}.
\end{eqnarray*}
Finally, by injecting  Eq.~\eqref{eq:computed-correlation-b} into Eq.~\eqref{eq:Gamma1-def}, we obtain the general expression for the longitudinal relaxation rate for a cylindrical cell immersed in a uniform gradient:
\begin{widetext}
\begin{eqnarray} \label{eq:gamma1-cylindricalcell-uniform}
    \Gamma _1= \frac{\gamma^2D}{\omega^2}\left[\left\vert \frac{\partial b}{\partial x} \right\vert^2 \ds\sum _{n=1,3,...} \frac{\left(\omega \frac{L^2}{D}\right)^2}{\left(\omega \frac{L^2}{D}\right)^2+ \left(n\pi\right) ^4}\frac{8}{(n\pi)^2}  + \left(\left\vert \frac{\partial b}{\partial y} \right\vert^2 + \left\vert \frac{\partial b}{\partial z} \right\vert^2\right) \ds\sum _{k\in\mathbb{N}^*}  \frac{\left(\omega \frac{R^2}{D}\right)^2 }{\left(\omega \frac{R^2}{D}\right)^2+ x_{1k} ^4}\frac{2}{x_{1k}^2-1} \right].
\end{eqnarray}
\end{widetext}

Now, this general expression can be simplified for specific shapes of the magnetic field. 
In particular, if $\frac{\partial b}{\partial x} = 0$ (valid in the experimental setup used in this study, see Eq.~\eqref{eq:field-shape}), Eq. \eqref{eq:gamma1-cylindricalcell-uniform} simplifies into Eq.~\eqref{eq:gamma1-final} given that 
\begin{equation}
    G^2 = \langle (\vec\nabla B_y)^2 + (\vec\nabla B_z)^2 \rangle =\left\vert \frac{\partial b}{\partial x} \right\vert^2 + \left\vert \frac{\partial b}{\partial y} \right\vert^2 + \left\vert \frac{\partial b}{\partial z} \right\vert^2.
\end{equation}



%

\end{document}